\documentclass[twoside,preprintnumbers,amsmath,amssymb,showpacs,twocolumn]{revtex4}
\usepackage{epsfig}
\usepackage{pslatex}
\usepackage{graphicx}
\usepackage{dcolumn}
\usepackage{bm}
\usepackage{braket}

\usepackage{amsthm} 

\newcommand{\p}{\partial}

\renewcommand{\d}{\ensuremath{\mathrm{d}}}

\begin{document}
\title{{\Large  New features of the gluon and ghost propagator in the infrared region from the Gribov-Zwanziger approach}}

\author{D. Dudal$^a$}
    \email{david.dudal@ugent.be}
    \author{S.P. Sorella$^b$}
    \email{sorella@uerj.br}
    \altaffiliation{Work supported by FAPERJ, Funda{\c c}{\~a}o de Amparo {\`a} Pesquisa do Estado do Rio de Janeiro,
                                    under the program {\it Cientista do Nosso Estado}, E-26/100.615/2007.}
\author{N. Vandersickel$^a$}
    \email{nele.vandersickel@ugent.be}
    \author{H. Verschelde$^a$}
 \email{henri.verschelde@ugent.be}
 \affiliation{\vskip 0.1cm
                            $^a$ Ghent University, Department of Mathematical Physics and Astronomy \\
                            Krijgslaan 281-S9, B-9000 Gent, Belgium\\\\\vskip 0.1cm
                            $^b$ Departamento de F\'{\i }sica Te\'{o}rica, Instituto de F\'{\i }sica, UERJ - Universidade do Estado do Rio de Janeiro\\
                            Rua S\~{a}o Francisco Xavier 524, 20550-013 Maracan\~{a}, Rio de Janeiro, Brasil}


\begin{abstract}
\noindent So far, the infrared behavior of the gluon and ghost
propagator based on the Gribov-Zwanziger approach predicted a
positivity violating gluon propagator vanishing at zero momentum,
and an infrared enhanced ghost propagator. However, recent data
based on huge lattices have revealed a positivity violating gluon
propagator which turns out to attain a finite \textit{nonvanishing}
value very close to zero momentum. At the same time the ghost
propagator does not seem to be infrared enhanced anymore. We point
out that these new features can be accounted for by yet unexploited
dynamical effects within the Gribov-Zwanziger approach, leading to
an infrared behavior in qualitatively good agreement with the new
data.
\end{abstract}

\maketitle

\setcounter{page}{1}

\section{Introduction.}
\noindent During the past few years, much attention has been devoted
to the study of the gluon and ghost propagator, including their low
energy behavior where Yang-Mills gauge theories are confining. As a
consequence the gluon cannot be considered as a free particle
anymore. Due to the lack of an explicit knowledge of the physical
degrees of freedom, it still remains highly useful to study the
gluon and ghost propagator in order to probe the nonperturbative
infrared regime. Let us only mention that the gluon propagator for
example finds an important use in phenomenological studies, see e.g.
\cite{Luna:2005nz}. In the past, a good agreement between the
lattice data and the analytical results arising from the
Gribov-Zwanziger action in the Landau gauge were found: (1) an
infrared suppressed and positivity violating gluon propagator
vanishing at zero momentum, (2) an infrared enhanced ghost
propagator. We recall that the Gribov-Zwanziger action was
constructed to take into account the existence of gauge copies
\cite{Gribov:1977wm,Zwanziger:1992qr}. However, very recent lattice
data obtained at large volumes
\cite{Cucchieri:2007md,Bogolubsky:2007ud}, which allows one to get
very close to zero momentum, now give evidence of a hitherto
unexpected behavior in the deep infrared: (1) an infrared suppressed
and positivity violating gluon propagator \textit{nonvanishing at
zero momentum}, (2) a ghost propagator essentially behaving like
$1/p^2$ at low momentum, which is clearly \textit{not enhanced}. To
our knowledge, none of the current theoretical approaches exhibit
all such features
\cite{Zwanziger:2001kw,Alkofer:2003jj,Pawlowski:2003hq,Aguilar:2004sw,Dudal:2005}.
In this letter we propose a dynamical mechanism within the
Gribov-Zwanziger approach that could account for the new lattice
results.

\section{The Gribov-Zwanziger action.}
\noindent We first give a short overview of the action constructed
by Zwanziger \cite{Zwanziger:1992qr} which implements the
restriction to the Gribov region $\Omega$ in Euclidean Yang-Mills
theories in the Landau gauge. We recall that this restriction to
$\Omega$ can be implemented by adding the nonlocal horizon function
to the original Yang-Mills action,
\begin{equation}\label{horizon}
S_{\mathrm{YM}}+ S_{\mathrm{Landau}} - \gamma ^{4}g^{2}\int
\d^{4}xf^{abc}A_{\mu }^{b}\left( \mathcal{M} ^{-1}\right)
^{ad}f^{dec}A_{\mu }^{e} \;,
\end{equation}
where $\mathcal{M}^{ab}=-\partial _{\mu }\left( \partial _{\mu
}\delta^{ab}+gf^{acb}A_{\mu }^{c}\right)$ is the Faddeev-Popov
operator, $S_{\mathrm{YM}} = 1/4 \int \d^4 x F_{\mu\nu}F_{\mu\nu}$
and $S_{\mathrm{Landau}}= \int \d^4 x (b^a \partial_{\mu}A^a_{\mu} +
\overline{c}^a \partial_{\mu} D^{ab}_{\mu} c^b)$ stands for the
gauge fixing and the ghost part. However, it is unclear how to
handle consistently such a nonlocal action at the quantum level, so
we are obliged to add extra fields
$\left(\overline{\varphi}_{\mu}^{ac},\varphi_{\mu}^{ac},
\overline{\omega}_{\mu }^{ac},\omega_{\mu}^{ac}\right)$ in order to
localize this action. Doing so, the Gribov-Zwanziger action reads
\cite{Zwanziger:1992qr,Dudal:2005}
\begin{equation}
S=S_{0}-\gamma ^{2}g\int \d^{4}x\left( f^{abc}A_{\mu }^{a}\varphi
_{\mu }^{bc}+f^{abc}A_{\mu }^{a}\overline{\varphi }_{\mu
}^{bc} + \frac{4}{g}\left( N^{2}-1\right) \gamma^{2} \right)\;,  \label{s1eq1}
\end{equation}
with
\begin{eqnarray}
S_{0} &=&S_{\mathrm{YM}}+\int \d^{4}x\;\left( b^{a}\partial_\mu A_\mu^{a}+\overline{c}%
^{a}\partial _{\mu }\left( D_{\mu }c\right) ^{a}\right) \;  \nonumber \\
&+&\int \d^{4}x\Bigl( \overline{\varphi }_{i}^{a}\partial _{\nu
}\left( D_{\nu }\varphi _{i}\right) ^{a}-\overline{\omega
}_{i}^{a}\partial _{\nu}\left( D_{\nu }\omega _{i}\right) ^{a}  \nonumber\\
&&   -g\left( \partial _{\nu }%
\overline{\omega }_{i}^{a}\right) f^{abm}\left( D_{\nu }c\right)
^{b}\varphi _{i}^{m}\Bigr)  \;,\label{s2eq4}
\end{eqnarray}
whereby $\left( \overline{\varphi }_{\mu }^{ac},\varphi
_{\mu}^{ac}\right) $ are a pair of complex conjugate bosonic fields,
whereas  $\left( \overline{\omega }_{\mu }^{ac},\omega
_{\mu}^{ac}\right) $ are anticommuting. Due to a global $U(f)$
symmetry, $f=4\left( N^{2}-1\right)$, with respect to the composite
index $i=\left( \mu ,c\right)$ of the additional fields $\left(
\overline{ \varphi }_{\mu }^{ac},\varphi _{\mu
}^{ac},\overline{\omega }_{\mu}^{ac},\omega _{\mu }^{ac}\right)$, we
introduced a notational shorthand,
\begin{equation}
\left( \overline{\varphi }_{\mu }^{ac},\varphi _{\mu }^{ac},\overline{\omega
}_{\mu }^{ac},\omega _{\mu }^{ac}\right) =\left( \overline{\varphi }%
_{i}^{a},\varphi _{i}^{a},\overline{\omega }_{i}^{a},\omega _{i}^{a}\right)
\;.  \label{l3}
\end{equation}
The dimensional parameter $\gamma$ is not free, being determined by
the gap equation (horizon condition)
 $\partial \Gamma / \partial \gamma = 0$, which ensures the restriction to the Gribov region. $\Gamma$ is the quantum effective action obtained from
 \eqref{s1eq1}.\\
 As it has been shown in \cite{Zwanziger:1992qr,Dudal:2005},
 the action  \eqref{s1eq1} is renormalizable to all orders. To prove this by the method
of algebraic renormalization \cite{Piguet:1995er}, we embed this
action into a larger one which contains more symmetries. This action
turns out to be given by \cite{Dudal:2005}
\begin{eqnarray}\label{cact}
\Sigma &=&S_{0}+S_{\mathrm{s}}+S_{\mathrm{ext}}\;,
\end{eqnarray}
with $S_{0}$ given in \eqref{s2eq4} and
\begin{eqnarray}
S_{\mathrm{s}} &=& s\int \d^{4}x\left( -U_{\mu }^{ai}\left( D_{\mu}\varphi _{i}\right) ^{a}-V_{\mu }^{ai}\left( D_{\mu
}\overline{\omega }_{i}\right) ^{a}-U_{\mu}^{ai}V_{\mu }^{ai} \right)\;,\nonumber\\
S_{\mathrm{ext}}&=&\int \d^{4}x\left( -K_{\mu }^{a}\left( D_{\mu }c\right)^{a}+\frac{1}{2}gL^{a}f^{abc}c^{b}c^{c}\right) \;.
\end{eqnarray}
We introduced new sources $M_{\mu}^{ai}$, $V_{\mu }^{ai}$
,$U_{\mu}^{ai}$, $N_{\mu}^{ai}$, $K_{\mu}^{a}$ and $L^{a}$, which
are necessary to analyze the renormalization of the corresponding
composite field operators in a BRST invariant fashion. The BRST
operator $s$ acts on the fields and sources appearing in the action
as follows
\begin{align}
sA_{\mu }^{a} &=-\left( D_{\mu }c\right) ^{a}\;,  sc^{a} =\frac{1}{2}gf^{abc}c^{b}c^{c}\;,  s\overline{c}^{a} =b^{a}\;, sb^{a}=0\;,  \nonumber \\
s\varphi _{i}^{a} &=\omega _{i}^{a}\;, s\omega _{i}^{a}=0\;,  s\overline{\omega }_{i}^{a} =\overline{\varphi }_{i}^{a}\;, s \overline{\varphi }_{i}^{a} =0\;,sU_{\mu }^{ai} = M_{\mu }^{ai}\;, \nonumber\\
 sM_{\mu }^{ai}&=0\;, sV_{\mu }^{ai} = N_{\mu }^{ai}\;,  sN_{\mu }^{ai} =0\;, s K_{\mu}^a = 0,  s L^{a} = 0\;,
\end{align}
whereby the BRST operator $s$ is nilpotent, $s^2 = 0$. One can
easily see that the action $\Sigma$ is indeed BRST invariant.
Henceforth, the action $\Sigma$ displays a greater number of
symmetries, encoded in the following Ward identities
\cite{Zwanziger:1992qr,Dudal:2005}.
\begin{itemize}
\item For the $U(f)$ invariance mentioned before we have
\begin{eqnarray}
U_{ij} \Sigma &=&0\;, \label{ward1} \\
U_{ij}&=&\int \d^{4}x\left( \varphi _{i}^{a}\frac{\delta }{\delta \varphi_{j}^{a}}-\overline{\varphi}_{j}^{a}\frac{\delta }{\delta \overline{\varphi}_{i}^{a}}+\omega _{i}^{a}\frac{\delta }{\delta \omega _{j}^{a}}-\overline{\omega }_{j}^{a}\frac{\delta }{\delta \overline{\omega}_{i}^{a}}\right) \nonumber\;.
\end{eqnarray}

\item  The Slavnov-Taylor identity reads
\begin{equation}
\mathcal{S}(\Sigma )=0\;,
\end{equation}
\begin{eqnarray}
\mathcal{S}(\Sigma ) &=&\int \d^{4}x\left( \frac{\delta \Sigma
}{\delta K_{\mu }^{a}}\frac{\delta \Sigma }{\delta A_{\mu
}^{a}}+\frac{\delta \Sigma }{\delta L^{a}}\frac{\delta \Sigma
}{\delta c^{a}}+b^{a}\frac{\delta \Sigma
}{\delta \overline{c}^{a}}+\overline{\varphi }_{i}^{a}\frac{\delta \Sigma }{%
\delta \overline{\omega }_{i}^{a}}\right.   \nonumber \\
&&\hphantom{\int \d^{4}x}+\omega _{i}^{a}\frac{\delta \Sigma }{\delta \varphi _{i}^{a}}+\left.M_{\mu }^{ai}\frac{\delta \Sigma
}{\delta U_{\mu}^{ai}}+N_{\mu }^{ai}\frac{\delta \Sigma }{\delta V_{\mu }^{ai}}\right) \;. \nonumber
\end{eqnarray}

\item  The Landau gauge condition and the antighost equation are given by
\begin{eqnarray}
\frac{\delta \Sigma }{\delta b^{a}}&=&\partial_\mu A_\mu^{a}\;,
\label{r11}\\ \frac{\delta \Sigma }{\delta
\overline{c}^{a}}+\partial _{\mu }\frac{\delta \Sigma }{\delta
K_{\mu }^{a}}&=&0\;.
\end{eqnarray}

\item  The ghost Ward identity:
\begin{eqnarray}
\mathcal{G}^{a}\Sigma &=&\Delta _{\mathrm{cl}}^{a}\;, \\
\mathcal{G}^{a} &=&\int \d^{4}x\left( \frac{\delta }{\delta c^{a}}+gf^{abc}\left( \overline{c}^{b}\frac{\delta }{\delta b^{c}}+\varphi _{i}^{b}\frac{\delta }{\delta \omega _{i}^{c}} \right.\right. \nonumber\\
&&\left.\left.+\overline{\omega }_{i}^{b}\frac{\delta }{\delta \overline{\varphi }_{i}^{c}} +V_{\mu }^{bi}\frac{\delta }{\delta N_{\mu }^{ci}}+U_{\mu }^{bi}\frac{\delta }{\delta M_{\mu }^{ci}}\right) \right) \;,  \nonumber \\
\Delta _{\mathrm{cl}}^{a}&=& g\int \d^{4}xf^{abc}\left( K_{\mu}^{b}A_{\mu }^{c}-L^{b}c^{c}\right) \;.  \nonumber
\end{eqnarray}
The term $\Delta _{\mathrm{cl}}^{a}$ denotes a classical breaking as
it is linear in the quantum fields.

\item  The linearly broken local constraints:
\begin{eqnarray}
&&\frac{\delta \Sigma }{\delta \overline{\varphi }^{ai}}+\partial _{\mu }\frac{\delta \Sigma }{\delta M_{\mu }^{ai}}=gf^{abc}A_{\mu }^{b}V_{\mu}^{ci}+J\varphi_i^a\;, \\
&&\frac{\delta \Sigma }{\delta \omega ^{ai}}+\partial _{\mu}\frac{\delta \Sigma }{\delta N_{\mu
}^{ai}}-gf^{abc}\overline{\omega }^{bi}\frac{\delta \Sigma }{\delta b^{c}}=gf^{abc}A_{\mu }^{b}U_{\mu }^{ci}+J\overline{\omega}_i^a\;, \\
&&\frac{\delta \Sigma }{\delta \overline{\omega }^{ai}}+\partial _{\mu }\frac{%
\delta \Sigma }{\delta U_{\mu }^{ai}}-gf^{abc}V_{\mu
}^{bi}\frac{\delta \Sigma }{\delta K_{\mu }^{c}}=-gf^{abc}A_{\mu
}^{b}N_{\mu }^{ci} -J\omega_i^a\;,\nonumber\\  \\
&&\frac{\delta \Sigma }{\delta \varphi ^{ai}}+\partial _{\mu}\frac{\delta \Sigma }{\delta V_{\mu
}^{ai}}-gf^{abc}\overline{\varphi }^{bi}\frac{\delta \Sigma }{\delta b^{c}}-gf^{abc}\overline{\omega }^{bi}\frac{\delta \Sigma }{\delta \overline{c}^{c}} \\
&& \hspace{2.5cm}-gf^{abc}U_{\mu }^{bi}\frac{\delta \Sigma}{\delta
K_{\mu }^{c}}  =gf^{abc}A_{\mu }^{b}M_{\mu
}^{ci}+J\overline{\varphi}_i^a\nonumber \;.
\end{eqnarray}

\item  The exact $\mathcal{R}_{ij}$ symmetry:
\begin{eqnarray}
\mathcal{R}_{ij}\Sigma &=&0\;,  \label{ward7}\\
\mathcal{R}_{ij}&=&\int \d^{4}x\left( \varphi
_{i}^{a}\frac{\delta}{\delta\omega _{j}^{a}}-\overline{\omega
}_{j}^{a}\frac{\delta }{\delta \overline{\varphi }_{i}^{a}}-V_{\mu
}^{ai}\frac{\delta }{\delta N_{\mu }^{aj}}+U_{\mu }^{aj}\frac{\delta
}{\delta M_{\mu }^{ai}}\right) \;.\nonumber
\end{eqnarray}
\end{itemize}
According to the algebraic renormalization procedure
\cite{Piguet:1995er}, Ward identities induce constraints on the most
general allowed counterterm $\Sigma^c$ at the quantum level. Once
$\Sigma^c$ is found, one can check if it can be reabsorbed in the
original starting action by a suitable renormalization of fields,
sources and parameters, thereby establishing the renormalizability.
One can show \cite{Dudal:2005} that $\Sigma ^{c}$ does not depend on
the Lagrange multiplier $b^{a}$, and that the antighost
$\overline{c}^{a}$ and the $i$-valued fields $\varphi_{i}^{a}$,
$\omega _{i}^{a}$, $\overline{\varphi }_{i}^{a}$, $\overline{\omega
}_{i}^{a}$ can enter only through the combinations
\begin{align}
&\widetilde{K}_{\mu }^{a} = K_{\mu }^{a}+\partial _{\mu }\overline{c}^{a}-gf^{abc}\widetilde{U}_{\mu }^{bi}\varphi ^{ci}-gf^{abc}V_{\mu }^{bi}\overline{\omega }^{ci}\;,  \nonumber \\
&\widetilde{U}_{\mu }^{ai} =U_{\mu }^{ai}+\partial _{\mu }\overline{\omega }^{ai}\;,
\widetilde{V}_{\mu }^{ai}  =V_{\mu }^{ai}+\partial _{\mu }\varphi^{ai}\;,\nonumber \\
&\widetilde{N}_{\mu }^{ai} =N_{\mu }^{ai}+\partial _{\mu }\omega^{ai}\;,
\widetilde{M}_{\mu }^{ai}  =V_{\mu }^{ai}+\partial _{\mu}\overline{\varphi }^{ai}\;.
\end{align}
Imposing the constraints, the most general counterterm yields,
\begin{eqnarray}
\Sigma ^{c}&=&a_{0}S_{YM}\nonumber\\
&& \hspace{-0.8cm}+a_{1}\int \d^{4}x\left( A_{\mu}^{a}\frac{\delta S_{YM}}{\delta A_{\mu}^{a}} +\widetilde{K}_{\mu }^{a}\partial _{\mu }c^{a}+ \widetilde{V}_{\mu }^{ai}\widetilde{M}_{\mu }^{ai} -\widetilde{U}_{\mu }^{ai} \widetilde{N}_{\mu }^{ai}\right)  \label{counterterm}
\end{eqnarray}
with $a_{0}$, $a_{1}$ two arbitrary parameters. It then turns out
that $\Sigma^c$ can be reabsorbed into the starting action
\eqref{cact} by a multiplicative renormalization
\cite{Zwanziger:1992qr,Dudal:2005}. At the end, we give the sources
the following physical values
\begin{eqnarray}
&\left. M_{\mu \nu }^{ab}\right|_{phys}= \left.V_{\mu \nu }^{ab}\right|_{phys}=\gamma ^{2}\delta ^{ab}\delta _{\mu \nu}\;,& \nonumber\\
&\left. U_{\mu }^{ai}\right|_{phys} = \left. N_{\mu }^{ai}\right|_{phys} =\left. K_{\mu }^{a}\right|_{phys} =\left. L^{a}\right|_{phys}  = 0\;,&
\end{eqnarray}
in order to recover the physical action \eqref{s1eq1}.

\section{Inclusion of a new dynamical effect.}
\noindent In a sense, the fields $\left( \overline{ \varphi }_{\mu
}^{ac},\varphi _{\mu }^{ac},\overline{\omega }_{\mu}^{ac},\omega
_{\mu }^{ac}\right)$ introduced to localize the horizon function
appearing in \eqref{horizon}, will correspond to the nonlocal
dynamics. Once these fields are present, they will quite evidently
develop their own dynamics at the quantum level, which might include
further nonperturbative effects, not yet accounted for. These
effects can induce important additional changes in the infrared
region. More precisely, looking at the $A\varphi$-coupling present
at tree level in the action \eqref{s1eq1}, a nontrivial effect in
the $\varphi$-sector will get immediately translated into the gluon
sector. We shall now explore the effects of a dynamical mass
generation for the $\varphi$-fields. This can be done by introducing
the local composite operator $\overline{\varphi}\varphi$ into the
action \eqref{s1eq1}. Since the horizon condition is in fact
equivalent with giving a particular value to a dimension 2
$A\varphi$-condensate \cite{Zwanziger:1992qr}, more precisely
$\langle
gf^{abc}A_\mu^a(\varphi_{\mu}^{bc}+\overline{\varphi}_{\mu}^{bc})\rangle=-2d\left(N^2-1\right)\gamma^2$,
it does seem to be reasonably fair to consider a possible
$\overline{\varphi}\varphi$-condensation. Remarkably, it turns out
that this is possible while preserving the renormalizability and
BRST invariance. In order to do so, we try to enlarge the action
$\Sigma$ by adding a massive term of the form
$J\overline{\varphi}^{a}_i \varphi^{a}_i$, with $J$ a new source.
First of all, for renormalization purposes, we have to add this term
in a BRST invariant way. Secondly, in analogy with
\cite{Dudal:2005}, we will also need a term $\propto J^2$,
indispensable to kill potential novel divergences $\propto J^2$ in
the generating functional. We thus consider the following extended
action:
\begin{eqnarray}
    \label{nact}\Sigma' &=& \Sigma + S_{\overline{\varphi} \varphi}, \\
    S_{\overline{\varphi} \varphi} &=& \int \d^4 x \left[s(-J \overline{\omega}^a_i \varphi^a_{i}) + \rho \frac{J^2}{2}\right]  \nonumber\\
    &=&\int \d^4 x \left[-J\left( \overline{\varphi}^a_i \varphi^a_{i} - \overline{\omega}^a_i \omega^a_i \right) + \rho \frac{J^2}{2}\right]   \;, \label{s3eq1}
\end{eqnarray}
with $\rho$ a new dimensionless quantity and $J$ a new source
invariant under the BRST transformation, $sJ = 0$. We underline that
the final mass operator,
$\overline{\varphi}\varphi-\overline{\omega}\omega$, is BRST
invariant. Now, it can be nicely checked that all the Ward
identities of the previous section are maintained. The final output
is that the new action \eqref{nact} enjoys multiplicative
renormalizability \cite{paperinprep}.\\ An interesting feature is
that the anomalous dimension of the mass $J$ is not an independent
quantity, as it is related to the running of the gauge coupling and
of the gluon field \cite{paperinprep}.
\\ We mention already that we will be able to prove that this new
parameter $\rho$ is in fact redundant. We postpone this to a
forthcoming larger paper \cite{paperinprep}, since the aim of this
letter is merely to illustrate the relevance of the introduced mass
operator. \\\\\noindent Summarizing, the BRST invariant mass
operator $\overline{\varphi}\varphi- \overline{\omega}\omega$ fits
quite naturally into the theory: it is renormalizable to all orders
and, moreover, it does not introduce any new renormalization
constants into the theory.

\section{The modified gluon and ghost propagator.}
\noindent Finally, we come to the main purpose of this letter. We
shall have a look at the effect on the propagators in the presence
of the mass operator $\overline{\varphi}\varphi-
\overline{\omega}\omega$.

\subsection{The gluon propagator.}
\noindent In order to calculate the gluon propagator we only need
the quadratic part of the action $\Sigma'$. We also replace the
source $J$ with the more conventional mass notation $M^2$, so that
\begin{eqnarray*}
    \Sigma'_0 &=& \int \d^4 x \Bigr( \frac{1}{4} \left( \p_{\mu} A_{\nu}^a - \p_{\nu}A_{\mu}^a \right)^2 + \frac{1}{2\alpha} \left( \p_{\mu} A^a_{\mu} \right)^2 + \overline{\varphi}^{ab}_{\mu} \p^2 \varphi^{ab}_{\mu} \nonumber\\
&&  - \gamma^2 g(f^{abc}A^a_{\mu} \varphi_{\mu}^{bc} +
f^{abc}A^a_{\mu} \overline{\varphi}^{bc}_{\mu} ) - M^2
\overline{\varphi}_{\mu}^{ab}\varphi_{\mu}^{ab}\Bigr) + \ldots  \;,
\end{eqnarray*}
where the limit $\alpha \rightarrow 0$ is understood in order to
recover the Landau gauge. The ``$\ldots$'' stand for the constant
term $\mbox{$-d (N^2 -1) \gamma^4$}$ and other terms in the ghost
and $\omega, \overline{\omega}$ fields irrelevant for the
calculation of the gluon propagator. Next, we integrate out
$\varphi$ and $\overline{\varphi}$, yielding
\begin{eqnarray*}
\Sigma'_0 &=& \int \d^4 x \frac{1}{2} A^a_{\mu} \Delta^{ab}_{\mu\nu} A^b_{\nu} + \ldots \;, \nonumber\\
\Delta^{ab}_{\mu\nu} &=&\left[ \left(-\p^2 - \frac{2 g^2 N
\gamma^4}{\p^2 - M^2} \right) \delta_{\mu\nu} + \p_{\mu}\p_{\nu}
\left(\frac{1}{\alpha} - 1\right) \right] \delta^{ab}.
\end{eqnarray*}
Taking the inverse of $\Delta^{ab}_{\mu\nu}$ and converting it into momentum space, we find the following gluon propagator
\begin{eqnarray}
 \mathcal{D}^{ab}_{\mu\nu} (p) &=&  \underbrace{ \frac{p^2 + M^2}{p^4 + M^2p^2 + 2 g^2 N \gamma^4 }}_{\mathcal{D}(p)}\mathcal{P}_{\mu\nu}(p)\delta^{ab} \;, \label{gluonprop}
\end{eqnarray}
where $\mathcal{P}_{\mu\nu}(p) =  \delta_{\mu\nu} -
\frac{p_{\mu}p_{\nu}}{p^2}$. From this expression we make three
observations: (1) $\mathcal{D}(p)$ enjoys infrared suppression, (2)
$\mathcal{D}(p)$ displays a positivity violation, (3)
$\mathcal{D}(0) \propto M^2$, so the gluon propagator does not
vanish at the origin, which is clearly a different result due to the
novel mass term proportional to $\overline{\varphi}\varphi-
\overline{\omega}\omega$.
\subsection{The ghost propagator.}
\begin{figure}[h]
   \centering
       \includegraphics[width=8cm]{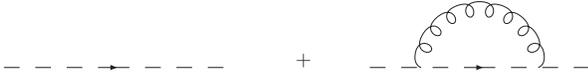}
   \label{ghost}
   \caption{The one loop corrected ghost propagator.}
\end{figure}
\noindent Let us now turn to the ghost propagator. We first derive
the gap equation for the Gribov parameter $\gamma$, useful for the
calculation of this propagator at one loop. The part of the one loop
effective action $\Gamma^{(1)}$ relevant for this gap equation reads
\begin{eqnarray*}
\Gamma_\gamma ^{(1)} &=& -d(N^{2}-1)\gamma^{4}\nonumber\\
&&+\frac{(N^{2}-1)}{2}\left( d-1\right) \int \frac{\d^{d}p}{\left(
2\pi \right) ^{d}}\ln \left( p^{4}+p^2 \frac{2Ng^{2}\gamma ^{4}}{p^2 + M^2}\right).
\end{eqnarray*}
Minimizing $\Gamma_\gamma^{(1)}$ with respect to $\gamma$ and
setting $d=4$ leads to the following gap equation:
\begin{eqnarray}\label{gapeq}
\frac{4}{3 g^2N } &=& \int \frac{\d^4 p}{(2\pi)^4} \frac{1}{p^4 +
M^2p^2 + 2g^2 N \gamma^4 }\;.
\end{eqnarray}
We are now ready to compute the ghost propagator at one loop order,
as depicted in FIG.1. The corresponding analytical representation
reads
\begin{eqnarray}
\mathcal{G}^{ab}(k) &=& \delta^{ab} \frac{1}{k^2} \frac{1}{1- \sigma}\;,
\end{eqnarray}
with
\begin{eqnarray*}
\sigma &=& Ng^2 \frac{k_{\mu} k_{\nu}}{k^2} \int \frac{\d^4
q}{(2\pi)^4} \frac{1}{(k-q)^2} \frac{q^2 + M^2}{q^4 + M^2 q^2 + 2
g^2 N \gamma^4}\mathcal{P}_{\mu\nu}(q) \;.
\end{eqnarray*}
If we take a closer look at the integral appearing in $\sigma$, we
can invoke the gap equation \eqref{gapeq} in order to simplify $(1 -
\sigma)$. By virtue of
\begin{eqnarray*}
&&Ng^2 \frac{k_{\mu} k_{\nu}}{k^2} \int \frac{\d^4 q}{(2\pi)^4} \frac{1}{q^4 + M^2 q^2 + 2 g^2 N \gamma^4}\left[ \delta_{\mu\nu} - \frac{q_{\mu}q_{\nu}}{q^2}\right]\nonumber\\
&&= Ng^2 \frac{k_{\mu} k_{\nu}}{k^2} \cdot \frac{3}{4} \delta_{\mu\nu} \int \frac{\d^4 q}{(2\pi)^4} \frac{1}{q^4 + M^2 q^2 + 2 g^2 N \gamma^4} = 1,
\end{eqnarray*}
we find
\begin{eqnarray}
1- \sigma &=& Ng^2 \frac{k_{\mu} k_{\nu}}{k^2} \int \frac{\d^4 q}{(2\pi)^4} \left[\frac{k^2-2k\cdot q - M^2}{(k-q)^2} \right] \nonumber\\
&& \qquad \times \frac{1}{q^4 + M^2 q^2 + 2 g^2 N \gamma^4}\mathcal{P}_{\mu\nu}(q) \;.
\end{eqnarray}
The last expression reveals that the ghost propagator will not be
enhanced at $k^2 = 0$. Indeed, if we expand $(1-\sigma)$ in the
region around $k^2 = 0$, we see
\begin{align}
& \hspace{-0.4cm}1-\sigma \nonumber\\
&\hspace{-0.4cm}= Ng^2 \frac{k_{\mu} k_{\nu}}{k^2} \int \frac{\d^4 q}{(2\pi)^4} \frac{- M^2}{q^2}  \frac{\mathcal{P}_{\mu\nu}(q)}{q^4 + M^2 q^2 + 2 g^2 N \gamma^4}  + \mathcal{O}(k^2) \nonumber\\
&\hspace{-0.4cm}= -\frac{3}{4}g^2 M^2 \frac{1}{8\pi^2} \frac{ \ln
\left( \frac{M^2 + \sqrt{M^4 - 8 g^2 N \gamma^4 }}{M^2 - \sqrt{M^4 -
8 g^2 N \gamma^4 } }\right) }{2 \sqrt{M^4 - 8  g^2 N \gamma^4 }}+
\mathcal{O}(k^2) \;.
\end{align}
We conclude that the ghost propagator keeps displaying a $1/k^2$
behavior for $k^2 \approx 0$. It becomes apparent now that, without
the introduction of the new BRST invariant mass term, the
Gribov-Zwanziger approach would predict a $1/k^4$ instead of a
$1/k^2$ behavior.

\section{Conclusion.}
\noindent In this letter, we have pointed out that the new lattice
data for the gluon and ghost propagator have a simple understanding
within the Gribov-Zwanziger approach. This is due to the
introduction of a multiplicatively renormalizable BRST invariant
mass operator $\overline{\varphi}\varphi- \overline{\omega}\omega$,
which fits into the theory in a very natural way. We hope that the
theoretical framework presented here will stimulate further
investigations, allowing a deeper understanding of the propagators
in the low momentum region. We end by noticing that it would be
interesting to find out what the analogous effects in the maximal
Abelian gauge might be and also compare those with available lattice
data \cite{Mendes:2006kc}.

\section{Acknowledgments.}
D.~Dudal is a Postdoctoral Fellow  and N.~Vandersickel a PhD Fellow
of the Research Foundation - Flanders (FWO).

\end{document}